\documentclass[journal=acsnano,manuscript=article,floatfix,10pt,superscriptaddress]{achemso}

\usepackage{bm}
\usepackage{graphicx}
\usepackage{subfigure}
\usepackage{epstopdf}
\usepackage{siunitx}
\usepackage{braket}
\usepackage{tabularx}
\usepackage{blindtext}
\usepackage{amsmath}
\usepackage{verbatim} 
\usepackage[version=3]{mhchem} 
\usepackage[symbol]{footmisc}
\usepackage{gensymb}
\usepackage{hyperref}

\usepackage{color}

\def\bem#1{\begin{mathletters}\label{#1}}
\def\eml{\end{mathletters}}

\def\4#1{{\boldsymbol{#1}}}
\def\8#1{{\widetilde{#1}}}

\title {Long-Timescale Magnetization Ordering Induced by an Adsorbed Chiral Monolayer on Ferromagnets}

\author{I. Meirzada} 
\affiliation{The Racah Institute of Physics, The Hebrew University of Jerusalem, Jerusalem 9190401, Israel}
\altaffiliation{Contributed equally to this work}

\author{N. Sukenik} \thanks{}
\affiliation{Dept. of Applied Physics, Rachel and Selim School of Engineering, Hebrew University, Jerusalem 9190401, Israel}
\affiliation{The Center for Nanoscience and Nanotechnology, The Hebrew University of Jerusalem, Jerusalem 9190401, Israel}
\altaffiliation{Contributed equally to this work}

\author{G. Haim} \thanks
\affiliation{Dept. of Applied Physics, Rachel and Selim School of Engineering, Hebrew University, Jerusalem 9190401, Israel}
\altaffiliation{Contributed equally to this work}

\author{S. Yochelis}
\affiliation{Dept. of Applied Physics, Rachel and Selim School of Engineering, Hebrew University, Jerusalem 9190401, Israel}

\author{L. T. Baczewski}
\affiliation{Magnetic Heterostructures Laboratory, Institute of Physics, Polish Academy of Sciences, Al. Lotnikow 32/46, 02-668 Warszawa, Poland}

\author{Y. Paltiel}
\affiliation{Dept. of Applied Physics, Rachel and Selim School of Engineering, Hebrew University, Jerusalem 9190401, Israel}
\altaffiliation{The Center for Nanoscience and Nanotechnology, The Hebrew University of Jerusalem, Jerusalem 9190401, Israel}

\author{N. Bar-Gill}
\affiliation{The Racah Institute of Physics, The Hebrew University of Jerusalem, Jerusalem 9190401, Israel}
\altaffiliation{Dept. of Applied Physics, Rachel and Selim School of Engineering, Hebrew University, Jerusalem 9190401, Israel}
\altaffiliation{The Center for Nanoscience and Nanotechnology, The Hebrew University of Jerusalem, Jerusalem 9190401, Israel}
\email{bargill@phys.huji.ac.il}

\begin{document}
\section{This document is the unedited author's version of a Submitted Work that was subsequently accepted for publication in ACS Nano, copyright © American Chemical Society after peer review. To access the final edited and published work, see  \url{https://pubs.acs.org/articlesonrequest/AOR-WCA48KFMFX5N4IM9SDVM}}
\begin{abstract}

When an electron passes through a chiral molecule there is a high probability for a correlation between the momentum and spin of the charge, thus leading to spin polarized current. This phenomenon is known as the chiral induced spin selectivity (CISS) effect. One of the most surprising experimental results recently demonstrated is that magnetization reversal in a ferromagnet (FM) with perpendicular anisotropy can be realized solely by chemisorbing a chiral molecular monolayer without applying any current or external magnetic field. This result raises the currently open question of whether this effect is due to the bonding event, held by the ferromagnet, or a long timescale effect stabilized by exchange interactions. In this work we have performed vectorial magnetic field measurements of the magnetization reorientation of a ferromagnetic layer exhibiting perpendicular anisotropy due to CISS using nitrogen-vacancy centers in diamond, and followed the time dynamics of this effect. In parallel, we have measured the molecular monolayer tilt angle in order to find a correlation between the time dependence of the magnetization re-orientation and the change of the tilt angle of the molecular monolayer. We have identified that changes in the magnetization direction correspond to changes of the molecular monolayer tilt angle, providing evidence for a long-timescale characteristic of the induced magnetization reorientation. This suggests that the CISS effect has an effect over long-timescales which we attribute to exchange interactions.
These results offer significant insights into the fundamental processes underlying the CISS effect, contributing to the implementation of CISS in state-of-the-art applications such as spintronic and magnetic memory devices.

\end{abstract}

{\bf Keywords:} CISS effect, self assembled monolayer, ferromagnetic thin films, NV centers, quantum magnetometry



\vspace{1cm}
Evidence of the Chiral Induced Spin Selectivity (CISS) effect have been measured for two decades\cite{ray_asymmetric_1999,Naaman2019}. 
When an electric charge is driven through a molecule with a certain chirality, spin selectivity is achieved - the charge that passes through the molecule has a preferred spin polarization. The same effect was demonstrated in many types of chiral molecules, including: double strand DNA\cite{ray_asymmetric_1999, gohler_spin_2011, abendroth_analyzing_2017}; Alpha-Helix or other types of protein that have both a helical structure and a chiral building block\cite{ben_dor_local_2014, kettner_spin_2015, ben_dor_magnetization_2017}; Helicene molecules that have only a helical geometry\cite{kiran_helicenesnew_2016}; small chiral molecules such as amino acids with only one chiral center\cite{bloom_spin_2016}. 
One of the most surprising phenomenon created by the CISS effect is the magnetization reorientation in a ferromagnetic layer with perpendicular anisotropy due to the chemical adsorption of chiral molecules of a given helicity on the thin film surface\cite{ben_dor_magnetization_2017, sukenik_correlation_2020}, while non-chiral molecules did not produce this effect. While many theories have been developed to address the basic properties of the CISS effect\cite{doi:10.1021/acs.jpcc.9b05020, medina_continuum_2015, gutierrez_spin-selective_2012, ghazaryan_analytic_2020, naaman_chiral_2020}, the underlying physical mechanism driving this effect is still unclear. One of the main open questions of interest, which is the focus of this work, is whether this effect is immediate - created only during the bonding process and results from it, or persistent - a steady state effect that is sustained over a long period of time, mediated by the exchange interaction. 

\begin{figure}[hbt!]
{\includegraphics[width = 0.9 \linewidth]{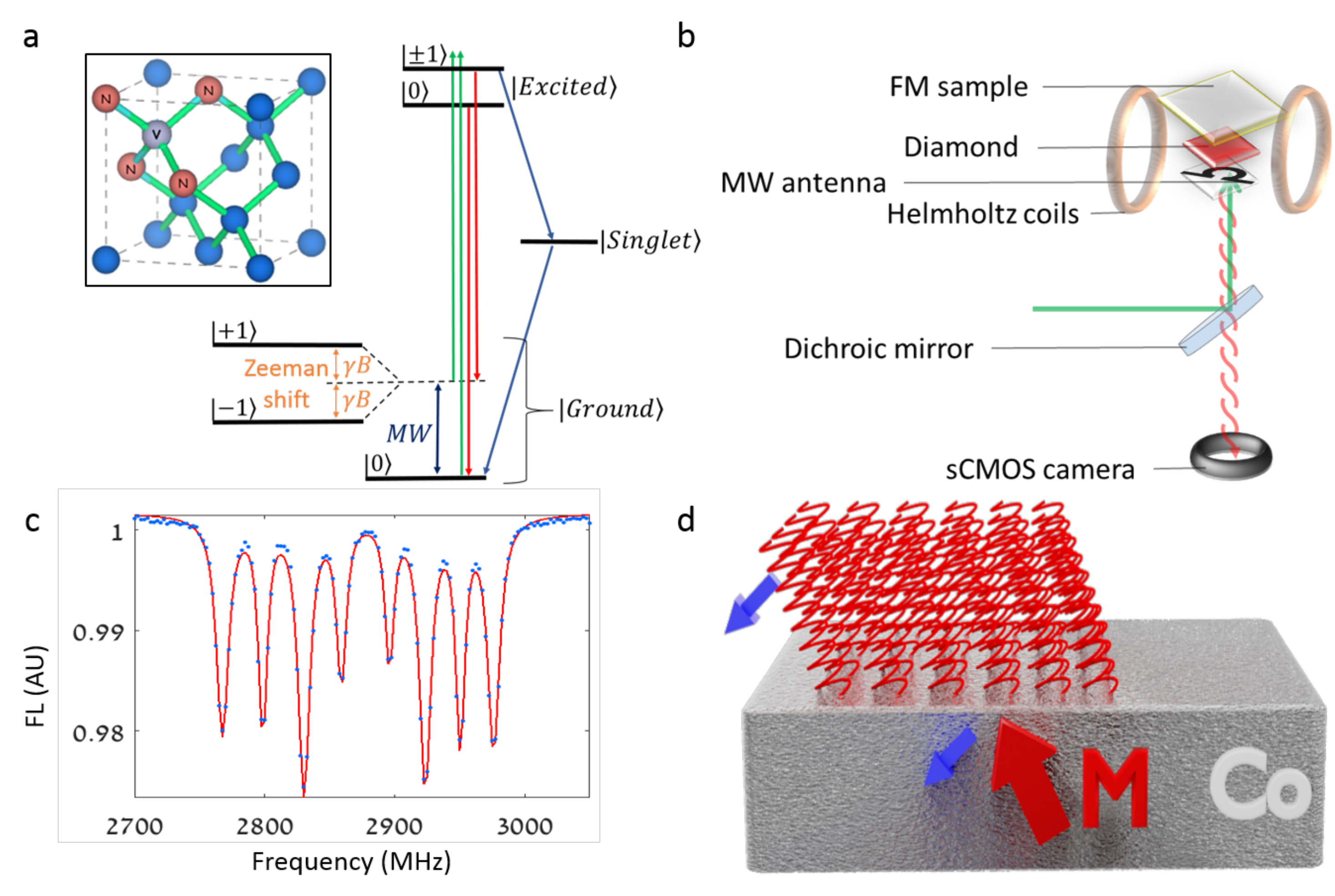}}
\caption{(a) NV center energy level diagram. Green light excites the electron from the ground states into the corresponding excited states. In zero field, the distance between $m_s = 0$ and $m_s = \pm 1$ is $2.87$ GHz. In the presence of an external magnetic field, Zeeman effect splits $m_s = \pm 1$ levels, the shift is proportional to $\gamma B$, where $\gamma= 2.8$  MHz/Gauss, is the gyromagnetic ratio of the NV. $m_s = 0$ is more likely to emit a photon (red arrows) once excited than $m_s = \pm 1$ due to significant coupling between $m_s = \pm 1$ to a non-radiative singlet state (blue arrow). inset: illustration of the four possible orientations an NV center can take in the diamond lattice. (b) Schematics of the NV wide field microscopy system. The sample is placed on top of the diamond, the adsorbed molecules layer facing a dense quasi 2D NV layer. The diamond is placed onto a fabricated $\Omega$-shaped waveguide, through which green laser excites the NV to the excited state, while red fluorescence is imaged onto an sCMOS camera after passing through a dichroic mirror. Helmholtz coils generate a magnetic field, splitting the energies of the four possible orientations in the diamond lattice, while the waveguide transfers the population between $m_s = 0$ and $m_s = \pm 1$. (c) Example of ESR measurement after splitting the four NV orientations. (d) Schematic illustration of the change in magnetization angle correlated with the change in molecular layer tilt angle as derived from the presented results}.  
\label{fig:NVsystem}
\end{figure}

The Nitrogen-Vacancy (NV) center \cite{doherty_nitrogen-vacancy_2013} has emerged as a promising system for a wide range of applications, from quantum information processing and control \cite{fuchs_quantum_2011}, to high precision sensing \cite{taylor_high-sensitivity_2008-2, loretz_nanoscale_2014-1, clevenson_broadband_2015-8, trusheim_wide-field_2016-6, dolde_electric-field_2011, farchi_quantitative_2017}. It is composed of adjacent nitrogen atom and vacancy, replacing two carbon atoms in the diamond lattice. The electronic ground state is a spin triplet, with a zero field splitting of 2.87 GHz between $m_s = 0$ and $m_s = \pm 1$ spin states, due to spin-spin interaction. Introducing an external magnetic field to the system causes the $m_s = \pm 1$ states energies to split due to the Zeeman effect (see Figure \ref{fig:NVsystem}b). The electronic excited state can decay radiatively to the ground state or non-radiatively to a metastable singlet manifold, with spin-dependent rates. The different rates and probabilities of these transitions make it possible to optically initialize and read the spin state, while spin manipulation, \emph{i.e.}, population transfer from  $m_s = 0$ to $m_s = \pm 1$, is possible using microwave (MW) fields (Figure \ref{fig:NVsystem}b).
The NV center has four possible orientations in the diamond lattice (Figure \ref{fig:NVsystem}b inset), along the crystallographic axes. In the presence of an external magnetic field, the spatial difference results in different Zeeman shift for each orientation, since the split is proportional to the field component parallel to the NV axis with correspondingly varied resonance frequencies. These resonances can be found by scanning MW frequencies and measuring the fluorescence intensity (Figure \ref{fig:NVsystem}c). Once the resonances are obtained, the vectorial sum will yield a vectorial quantitative field measurement.
\\Due to their optical and quantum properties\cite{bar-gill_solid-state_2013-3,farfurnik_optimizing_2015,childress_coherent_2006,robledo_high-fidelity_2011}, NV centers excel in vectorial magnetic field imaging\cite{tetienne_magnetic-field-dependent_2012-1,RevModPhys.92.015004}, demonstrating sub-micron resolution and sensitivities of up to $\sim \frac{picoTesla}{\sqrt{Hz}} $, depending on sensor area (or volume)\cite{zheng_zero-field_2019,wolf_subpicotesla_2015,balasubramanian_dc_2019}. 

In this work we employ NV wide-field microscopy to follow the dynamical behavior of magnetization properties of thin ferromagnetic layers with perpendicular anisotropy, after adsorption of Alpha-Helix L polyalanine (AHPA) chiral molecules on their surface. These measurements provide a quantitative analysis of the magnetization angle and magnitude, allowing the distinction between different potential mechanisms and between the immediate or persistent nature of CISS. 

In order to correlate these magnetic results with the behavior of the molecular monolayer, atomic force microscopy (AFM) measurements were performed to measure changes in the topography of the monolayer over time, as a direct measurement of the tilt angle of the molecules. The presented results indicate that the magnitude of the adsorption induced magnetization reorientation decays over time. Furthermore, the trend in magnetization tilt angle corresponds to the trend of the molecular monolayer tilt angle, as illustrated in Figure \ref{fig:NVsystem}d, thus providing evidence for a persistent steady-state effect rather than an immediate one. 

\section{Results and Discussion}

\begin{figure}[tbh]
{\includegraphics[width = 0.7 \linewidth]{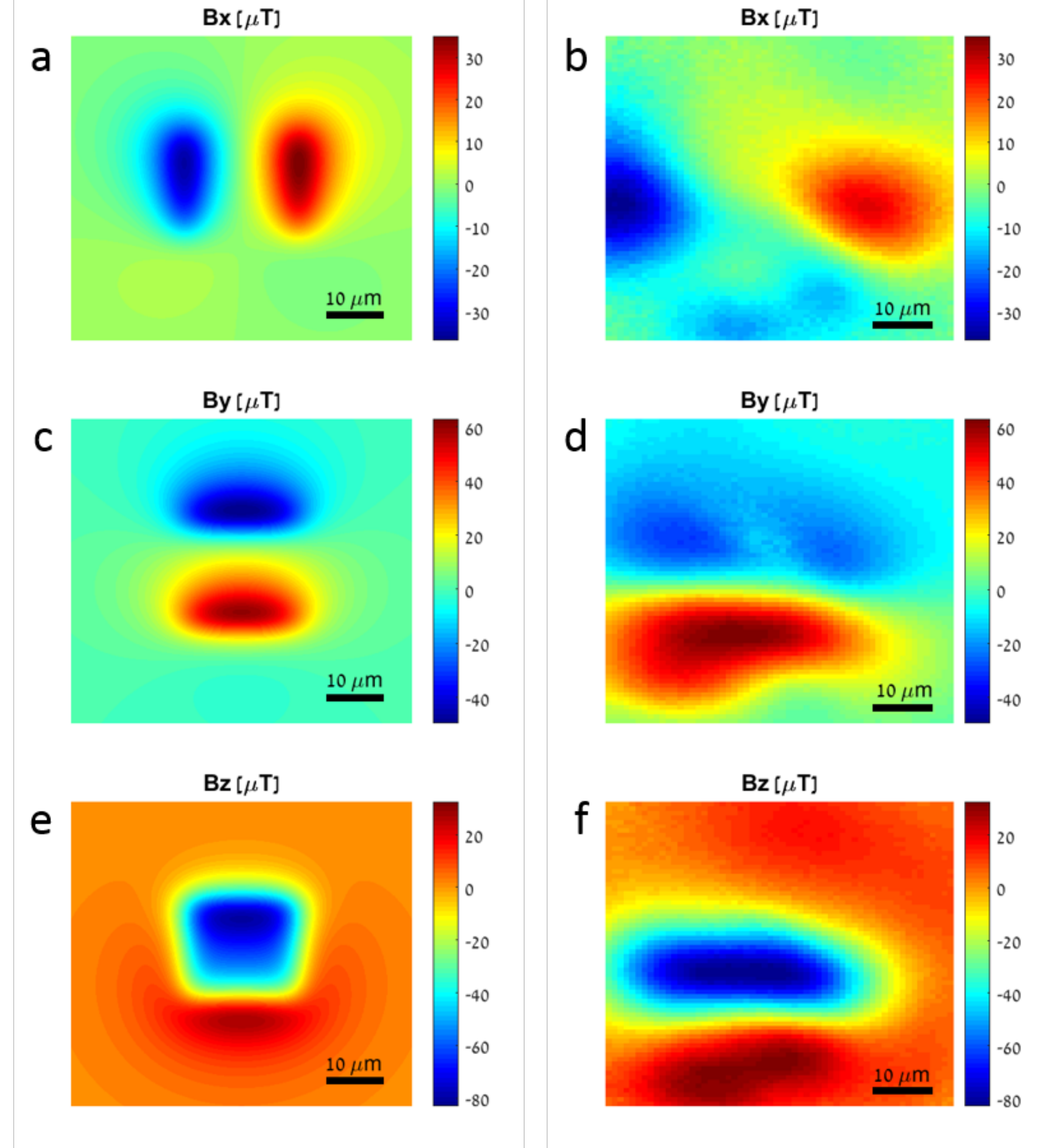}}
\caption{Vectorial magnetization measurement and simulation. Magnetization recorded by the NV system along with X (a), Y (c), and Z (e) directions, together with their corresponding simulations (b,d,f).  } 
\label{fig:ssmagnetization}
\end{figure}

In Figure \ref{fig:ssmagnetization}(a,c,e) we present the vectorial stray magnetic field generated by the ferromagnet measured a few hours after adsorption in X, Y and Z directions, correspondingly. These measurements were performed at a stand-off distance of $\sim$ 10 $\mu$m between the NV-containing diamond and the sample, for which the dipole approximation of the stray field holds. The results are consistent with an averaged magnetic field generated by a dipole structure corresponding to the adsorption pattern (see methods section) at a distance of few $\mu$ms, as further explained in the SI. 
The corresponding simulations, depicted in Figure \ref{fig:ssmagnetization}(b,d,f), optimize the tilt ("$\theta$") and azimuthal ("$\phi$") angles (further defined in the SI)  of the spins to reproduce the shape and ratio between the positive and negative regions of the magnetic field image, assuming equal magnetization in all adsorbed regions (both magnitude and angle). The simulation indicates that the magnetization tilt angle was 40 $\pm$ 10 degrees with respect to the sample normal, and the azimuthal angle was 90 $\pm$ 3 degrees (see SI). The extracted $\phi$ angle describes the average azimuthal angle throughout the adsorption area. Although theoretically different adsorption areas may have different azimuthal angles, these cannot be random as this would diminish the in-plane magnetic field, manifested in the secondary (red) "blob" in the Z component of the magnetic field (see SI). Thus, the existence of a measurable field at this second "blob" strongly suggests that the azimuthal angles of the magnetic dipoles are distributed around the extracted value.

Several uncertainties regarding the experimental conditions can affect the precise simulation of the results. These include: variations in the stand-off distance, inhomogeneities in the NV strain properties, and potential variations in the azimuthal $\phi$ angle of the adsorbed molecules within the measured region. Such uncertainties can affect the direct correspondence between the measurement and simulation results. 
However, it is important to emphasize that once the azimuthal $\phi$ angle is known, the ratio between the positive and negative “blobs” in the X, Y and Z components of the magnetic field depends solely on the tilt angle of the molecules and on the standoff distance between the diamond and the sample, while the shape, the distance between the “blobs” and additional structural characteristics are affected by other factors (such as the $\phi$ angle and actual adsorption area and shape). Thus, our conclusions regarding the tilt angle of the molecules is robustly obtained from the sign and magnitude of the different “blobs” with the stated accuracy, while the comparison of other properties of the simulation and measurement results need to be examined from a qualitative perspective.

We note that the quantitative measurement with the NV microscope enables us to estimate the fraction of polarized electrons in the FM layer. In order to simulate the measured magnitude of the magnetic field shown in Figure \ref{fig:ssmagnetization}, at least 10$\%$ of the cobalt electrons have to be polarized. Since there is about three orders of magnitude more free electrons in the Co than adsorbed molecules on the surface, this result suggests a strong coupling between the molecules and the cobalt electrons, and presents an important additional insight into the fundamental nature of this phenomenon.

\begin{figure}[tbh]
{\includegraphics[width = 1\linewidth]{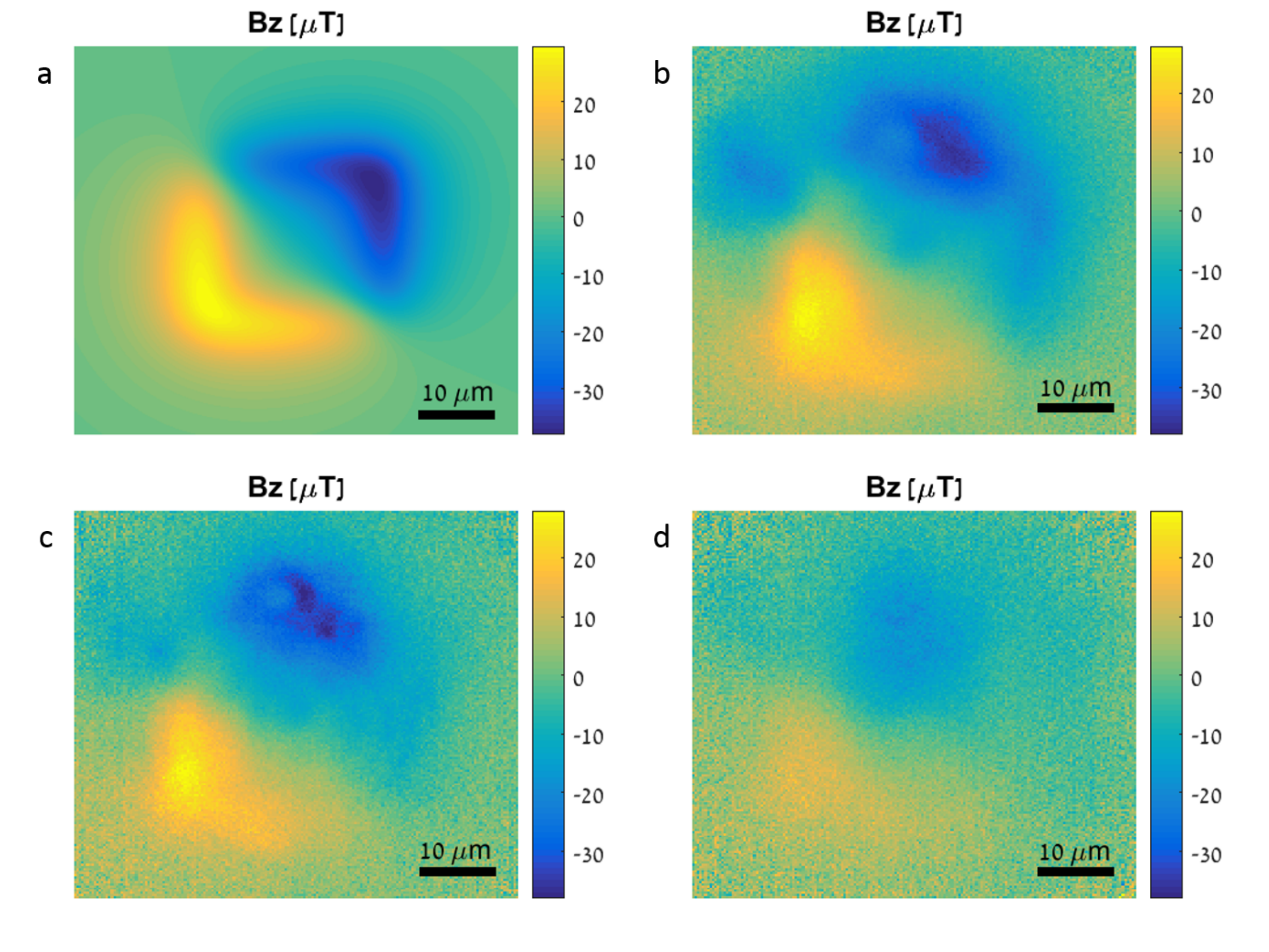}}
\caption{(a) Simulation of the magnetization along the Z axis at time $t = 4$ hours (following adsorption), corresponding to angles $\phi$ = 120 $\deg$ and $\theta$ = 80 $\deg$. (b-d) Measured magnetization along the Z direction, recorded using the NV microscope at times $t=4$ (b), $t=8$ (c) and $t=12$ hours (d) after adsorption, corresponding to angles $\theta = 80$, $\theta = 84$ and $\theta = 86.5$ degrees.} 
\label{fig:Bztime}
\end{figure}
\begin{figure}[tbh]
{\includegraphics[width = 1 \linewidth]{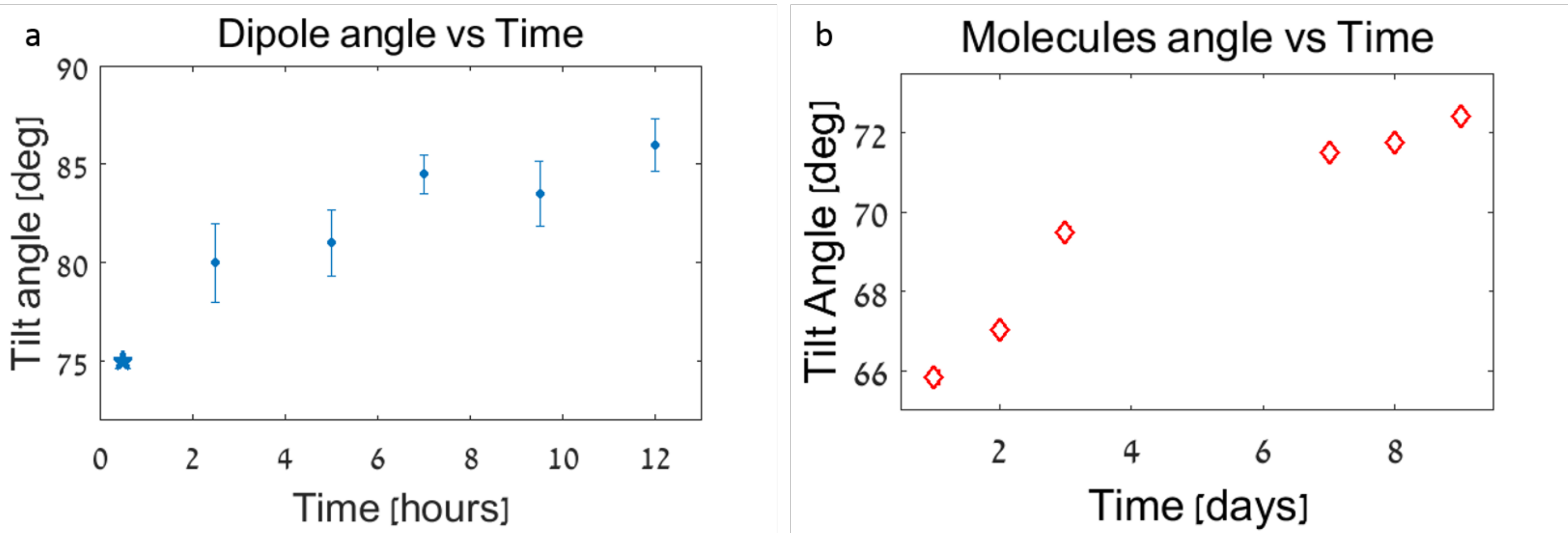}}
\caption{Change in the ferromagnet magnetization tilt angle and the tilt angle of the molecular layer as a function of time. (a) The tilt angle of the magnetization in the Co layer, extracted from the stray magnetic field measurements of the NV system and their comparison with an effective magnetic dipole representing the magnetization in the FM, as described in figure \ref{fig:Bztime} (for the same batch of molecules). The blue star depicts an initial AFM measurement of the height of the molecular layer, from which the tilt angle of the molecules is extracted, performed on the same sample before the NV measurements. (b) The measured tilt angle of the molecular monolayer for a different batch of the same molecules (AHPA) on a similar FM substrate, as was extracted by the height of the molecules using an AFM measurement.} 
\label{fig:time_summary}
\end{figure}

As a function of time, the magnitude of the magnetic field in the Z direction decreases as the magnetization tilt angle increases (relative to the surface normal of the sample). Figure \ref{fig:Bztime} follows the average magnetization along the Z axis measured at $t=4$, $t=8$ and $t=12$ hours after adsorption, revealing a significant weakening over time. The ratio between magnitudes of the magnetic "blobs" reflects the angle of the magnetization vector. During the measurement this ratio changes as well, becoming more balanced, indicating that the dipole direction 'falls' into the sample plane as a function of time. 
Figure \ref{fig:time_summary}.a summarizes the magnetic dipole angle extracted by these measurements as a function of time. This was supported by a topographic AFM measurement of the same batch of molecules on a similar magnetic substrate immediately after the adsorption process, yielding a tilted monolayer of roughly 75 degrees with respect to the normal, depicted as a blue star.

Next, we have repeated the adsorption process using a different batch of molecules and measured their height from the sample surface as a function of time, using an AFM tip. The measured height directly provides the tilt angle of the molecular layer with respect to the surface normal, based on the calculated length of the AHPA molecules ($\sim$ 5nm). A similar tilt angle was measured using Fourier transform infra-red spectroscopy (FTIR) (see SI). The results of these measurements, summarized in Figure \ref{fig:time_summary}b, show that the adsorption tilt angle increases with time, agreeing with the trend in magnetization angle as a function of time recorded by the NV system. This agreement suggests that the magnetization direction in the FM follows the angle of the molecular monolayer with respect to the FM layer normal, consistently over hours, as this angle changes. This correspondence provides direct evidence to the persistent nature of the CISS effect, constituting the main result of this paper. We note that the initial tilt angle is different between the two batches. However, we attribute this to inherent randomness in the sample preparation process. 

The above presented results show a direct correlation between the magnetization angle of the FM layer and the bonding angle of the molecules. Previous measurements, performed both by AFM/MFM\cite{ben_dor_magnetization_2017,sukenik_correlation_2020} and Magneto optical Kerr Effect (MOKE) microscopy\cite{sharma_control_2020}, could not provide quantitative, vectorial stray magnetic field information, and thus could only show that chemisorbing chiral molecules on a FM can reorder the magnetization and even flip it. Therefore previous results were not sensitive to magnetization angle and magnitude, and could not rule out the possibility that the magnetization re-orientation is induced solely during the adsorption process due to transient charge redistribution. 

The data presented here shows that the interaction between the chiral molecules and the magnetic Co structure is not only a result of the electrochemical changes that occur during the relatively fast, transient adsorption process, but rather strong enough to remain stable over long periods of time (hours to days). Adsorption of molecules that polarize only 10\% of the Co electrons is enough to tilt over time (hours) the Co layer easy axis (perpendicular to the layer due to anisotropy), following the angle change of the molecules. 

Motivated by these results, a theoretical approach explaining this CISS induced magnetization phenomena was suggested\cite{naaman_chiral_2020}. This theory claims that charge polarization, caused due to a change in the electric field or electrochemical potential, is accompanied by spin polarization along the chiral molecule. This spin polarization acts as a 'magnetic dipole' in the molecule, which splits the energies of electrons (or holes) penetrating the molecule, depending on their spin state (spin parallel or anti parallel to the 'magnetic dipole' of the molecule). This energy splitting leads to a spin blockade effect, which translates to an induced magnetization change in the FM layer. 
The strength of the exchange interaction between the molecules and the substrate should be larger than $10 \times K_B T$ (since only 10\% of the Co electrons are coupled), \emph{i.e.} larger than 250 meV. This energy range was predicted recently\cite{Gianaurelio2020} and estimated experimentally\cite{Paltiel2019_AdvMat}.
Therefore this theory\cite{naaman_chiral_2020} is consistent with the experimental results presented in this work, since a change in molecular bond angle (that could be caused by several mechanisms such as changes in hydrogen bonds between the molecules, Sulphurization of the substrate, electrostatic attraction/repulsion between the molecular backbone/carboxylic groups \emph{etc.}) modifies the molecule's effective magnetic dipole moment, thus changing the magnetization direction in the FM.

During measurements of different samples, various initial magnetization and adsorption angles were obtained by both the AFM and the NV system, ranging from 40 to 75 degrees. We relate the different initial tilt angles to varied densities of the adsorbed molecular layer \cite{gatto_electrochemistry_2016}, since this density could affect the tilt angle due to molecular crowding. Both our NV and AFM measurements support this hypothesis, as they recorded different changes in the magnetization and tilt angles for different initial tilt angles in two different samples.

\section{Conclusions}
To summarize, we present quantitative, vectorial measurements of magnetization reorientation induced by chiral molecules on a thin FM layer with perpendicular anisotropy, performed by NV-based magnetic microscopy. These measurements, correlated with AFM measurements of the molecular monolayer height, attest that the magnetization direction follows the adsorption angle of the molecules, and changes slowly toward the in-plane direction over the course of hours. This indicates that the CISS-induced magnetization effect is persistent and not immediate, resolving a long time open question in the field. This effect stems from coupling between a magnetic substrate and the magnetic dipole across the chiral molecule layer, which is stabilized over time through large spin exchange energy\cite{Gianaurelio2020,Paltiel2019_AdvMat}.

The NV-based measurement modality presented here has the potential to enable significant additional measurements of this and related effects, such as magnetization and coherent spin dynamics down to the few/single spin level. As such we expect that this tool could contribute further to the understanding of thin magnetic films and their magnetic exchange coupling to chiral nanostructures, quantum properties of the induced spin dynamics, and more.

\section{Methods}

\begin{figure}[tbh]
{\includegraphics[width = 0.9 \linewidth]{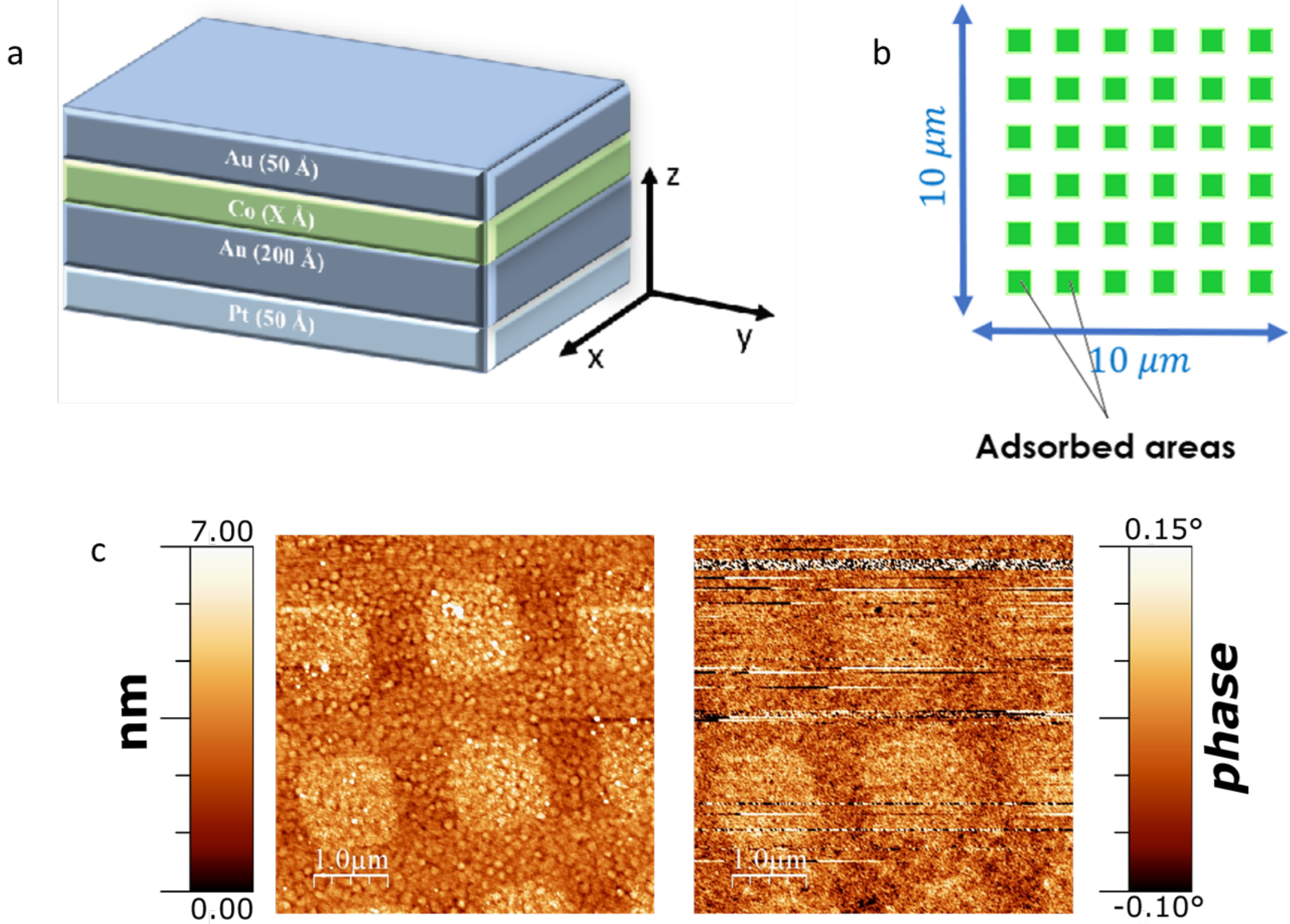}}
\caption{Schematics of the measured samples. a. Illustration of the Ferro-magnetic substrate with different Co thickness. b. Schematics of the pattern of the adsorbed areas. c. Topography of the adsorbed pattern using AFM tip in tapping mode, for which the dense molecular monolayer appears continuous and $\sim$ 2 nm height (left), and magnetization of the adsorbed pattern using magnetic AFM tip (right). The adsorbed areas clearly show magnetic response, despite some noise caused by contamination on the sample surface.} 
\label{fig:adsorption}
\end{figure}

Epitaxial thin film samples of varying cobalt (Co) thickness (1.6 - 1.8 nm) were grown by Molecular Beam Epitaxy (PREVAC MBE) as illustrated in Figure \ref{fig:adsorption}a. All layers of the sample are continuous and homogeneous since they are much thicker than an atomic monolayer.
The Co layer is capped on both sides with gold (Au), which ensures Co epitaxial growth, prevents oxidation and enables covalent bonding to the chiral molecules. Furthermore, the Au layer enhances the perpendicular anisotropy in Co film what enables magnetization reorientation due to chemically bonded chiral molecules. 
The samples magnetization easy axis is out of plane, with a coercive field of $\sim 150 G$
The molecules were adsorbed selectively in a checkerboard pattern, drawn in Figure \ref{fig:adsorption}b, containing 6 by 6 squares of  1x1 $\mu m^2$ areas, fabricated in PMMA using E-beam lithography.
The molecules used in this experiment were alpha helix poly alanine (AHPA - 36 amino acids, see SI for full molecular structure) with a calculated length of 5.4 nm. A self-assembled monolayer (SAM) was then deposited onto the patterned areas by dipping in a 1 mM  Ethanol solution of AHPA for 3 hours. The PMMA was then removed by rinsing the sample with Acetone, leaving only the covalently bonded molecules in the adsorbed areas, followed by rinsing in Ethanol and drying in Nitrogen. Figure \ref{fig:adsorption}c depicts representative AFM topography and MFM measurements of the adsorbed areas.  

The NV experimental system, schematically illustrated in Figure \ref{fig:NVsystem}a, consists of a custom-built Wide-field magnetic imaging epi-illumination microscope. The diamond is placed on a fabricated $\Omega$-shaped MW antenna, used for spin manipulation.
A green laser (Laser Quantum Ventus 532 nm, 1 W), used for initialization and read out, is directed onto the diamond through an air objective (CFI Plan Apo Lambda 60XC), and an sCMOS camera (Andor Neo 5.5) collects the red NV fluorescence following a dichroic mirror (Semrock Di03-R561-t1-25x36). 
Helmholtz coils generate a controlled bias magnetic field $(\sim 50 G$, well below the Co coercive field) for splitting the electron spin resonance (ESR) spectra of all four NV orientations (Figure \ref{fig:NVsystem}c), enabling a CW, 16-point ESR measurement\cite{barry_sensitivity_2020} per pixel, from which a 2D vectorial magnetic image can be extracted. 

The ferromagnetic sample is placed on the diamond, such that the facet with adsorbed molecules faces the NV layer as shown in Figure \ref{fig:NVsystem}b, resulting in a standoff distance of $\sim 10 \mu$m.

\section{Acknowledgements}
Y.P. acknowledges support from the Ministry of Science and Technology, Israel. 
N.B. acknowledges support from the Ministry of Science and Technology, Israel, and the European Union’s Horizon 2020 research and innovation program under grant agreements No. 714005 (ERC StG Q-DIM-SIM), No. 820374 (MetaboliQs), and No. 828946 (PATHOS).
G.H and N.S thank Dr. Yael Ebert and Prof. Ron Shaar for assisting with samples preparation. 

\section{Supporting Information}

The Supporting Information Available online 
Details of the NV microscopy calibration and magnetization analysis, information regarding the magnetic characterization of the FM sample, Alpha Helix L-Polyalanine specifications, supporting FTIR measurements and AFM height analysis. 


\bibliography{ChiralMolecules.bib}

\end{document}


\noindent $^\dag$ The Racah Institute of Physics, The Hebrew University of Jerusalem, Jerusalem 9190401, Israel
\newline
$^\ddag$ Dept. of Applied Physics, Rachel and Selim School of Engineering, Hebrew University, Jerusalem 9190401, Israel
\newline
$^{\S}$ The Center for Nanoscience and Nanotechnology, The Hebrew University of Jerusalem, Jerusalem 9190401, Israel
\newline
$^\parallel$ Magnetic Heterostructures Laboratory, Institute of Physics, Polish Academy of Sciences, Al. Lotnikow 32/46, 02-668 Warszawa, Poland
\newpage
\section{NV microscope calibration}
The magnetic measurements described in the main text were performed using a home-built optical magnetic microscope using NV centers in diamond. The system is detailed in the Methods section of the main text.
\newline Here we detail the calibration procedures, both optical and magnetic, performed on the system to analyze its sensitivity and resolution.
\subsection{Magnetic calibration}
Our magnetic measurement is calibration-free, as the NVs measure absolute vectorial magnetic field. 
Nevertheless, we verify this and compare the magnetic field created by our Helmholtz coils using both the NV system and a separate Gaussmeter placed at the same location (Lakeshore 425). The magnetic image measured by the NV system is depicted in Figure \ref{fig:calibration}(a), and is consistent with the field measured by the Gaussmeter. Moreover, the image indicates a magnetic field homogeneity of $< 1 \mu$T over the entire field-of-view.

We note this system was used in previous magnetic measurements performed in the context of Paleomagnetism \cite{farchi_quantitative_2017}, and are similar to magnetic measurements carried out in equivalent setups \cite{taylor_high-sensitivity_2008-2,trusheim_wide-field_2016-6,tetienne_apparent_2019, FU20171, 2003.08470, 10.1130/G39938.1, hall_high_2012, Tetiennee1602429}.

\subsection{Optical calibration}
\begin{figure}[hbt!]
{\includegraphics[width = 0.9 \linewidth]{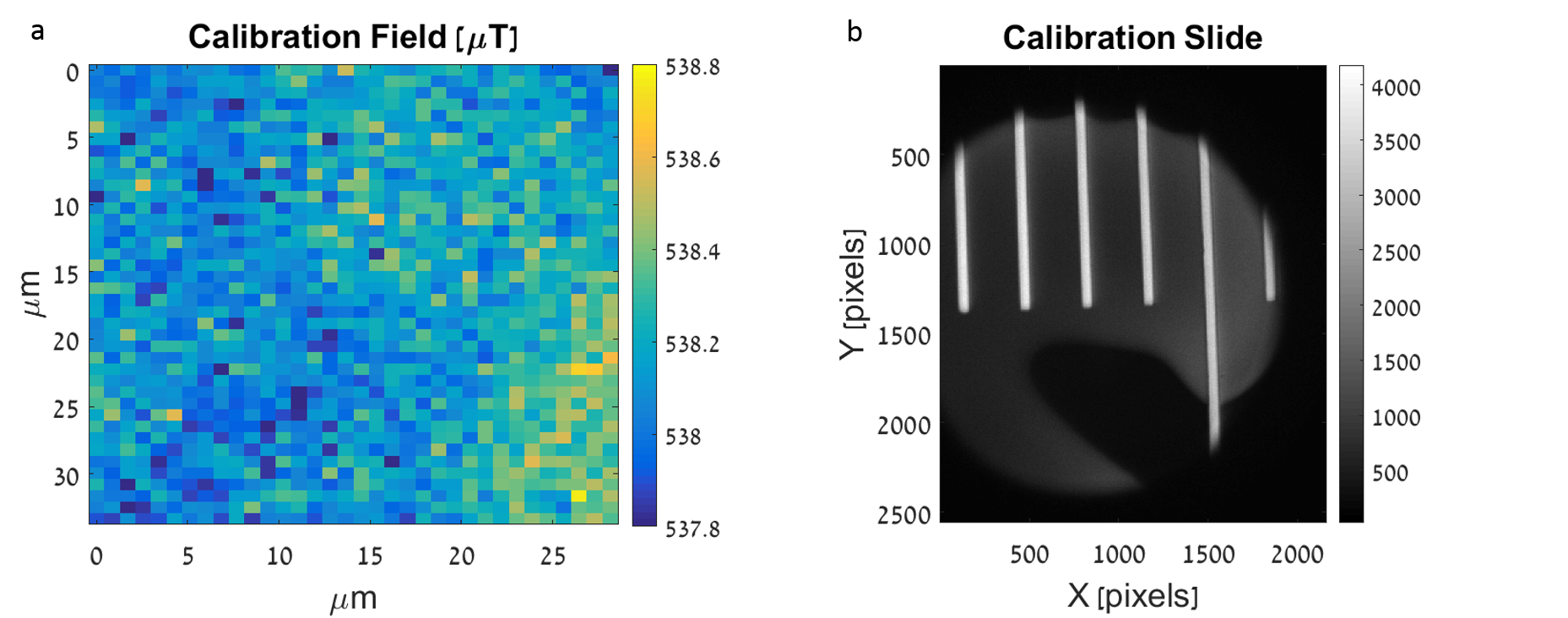}}
\caption{Calibration measurements: (a) Magnetic imaging of the field produced by the Helmholtz coils was compared to a separate measurement using a Gaussmeter at the same location.} (b) Optical calibration of the imaging system performed using a calibration slide, determining the pixel size and microscope amplification.
\label{fig:calibration}
\end{figure}

The optical image is recorded by an sCMOS camera (Andor NEO 5.5). We have calibrated our imaging system and extracted the pixel resolution using a standard calibration slide (lines separated by $10 \mu$m), as illustrated in figure \ref{fig:calibration}b. 

\section{Vectorial magnetization vs dipole orientation}
\begin{figure}[hbt!]
\subfigure[]
{\includegraphics[width = 0.32 \linewidth]{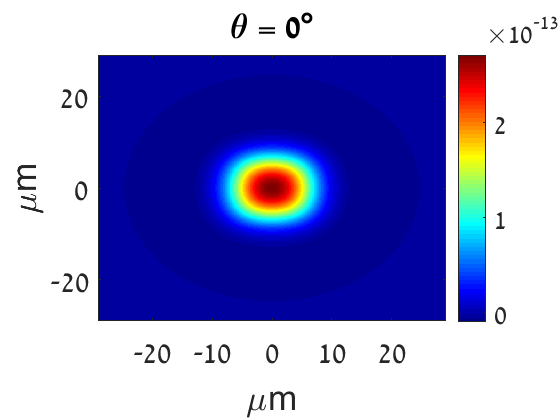}}
\subfigure[]
{\includegraphics[width = 0.32 \linewidth]{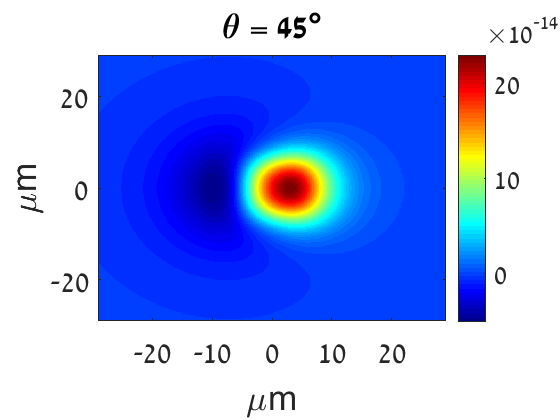}}
\subfigure[]
{\includegraphics[width = 0.32 \linewidth]{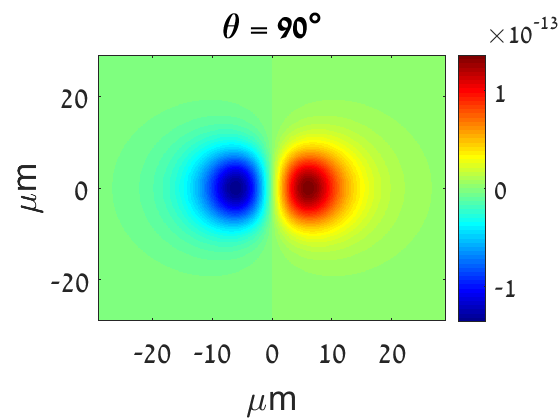}}
\caption{Simulation of the vectorial magnetization (Arb. units) in the Z direction as a function of dipole angle $\theta$, from 0 (perpendicular to sample plane) to 90 degrees from a distance of 7 $\mu$m from the sample. Azimuth angle $\phi$ = 0}. 
\label{fig:analysis_theta_z}
\end{figure}

\begin{figure}[hbt!]
\subfigure[]
{\includegraphics[width = 0.32 \linewidth]{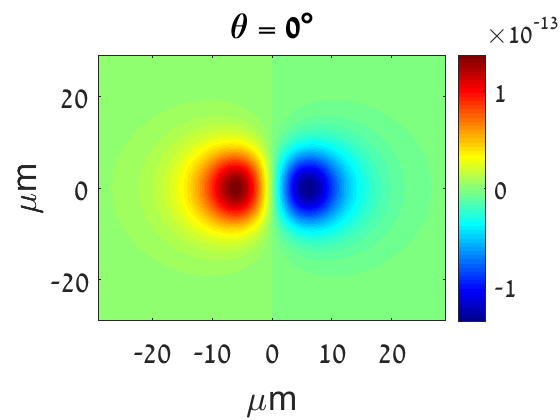}}
\subfigure[]
{\includegraphics[width = 0.32 \linewidth]{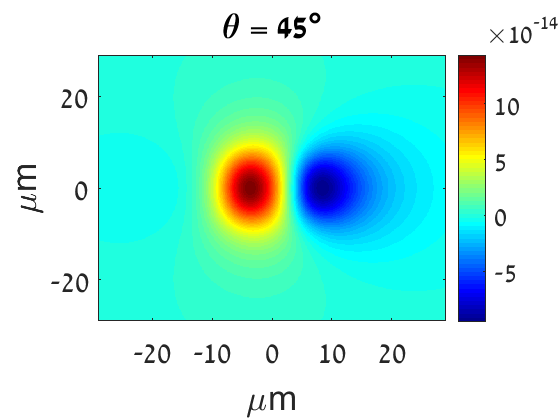}}
\subfigure[]
{\includegraphics[width = 0.32 \linewidth]{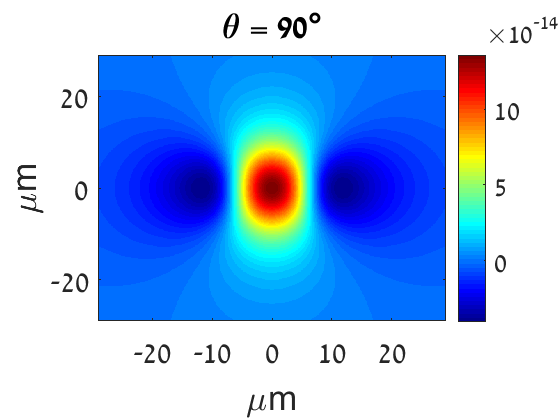}}
\caption{Simulation of the vectorial magnetization (Arb. units) in the X direction as a function of dipole angle $\theta$, from 0 (perpendicular to sample plane) to 90 degrees from a distance of 7 $\mu$m from the sample. Azimuth angle $\phi$ = 0}. 
\label{fig:analysis_theta_x}
\end{figure}

\begin{figure}[hbt!]
\subfigure[]
{\includegraphics[width = 0.32 \linewidth]{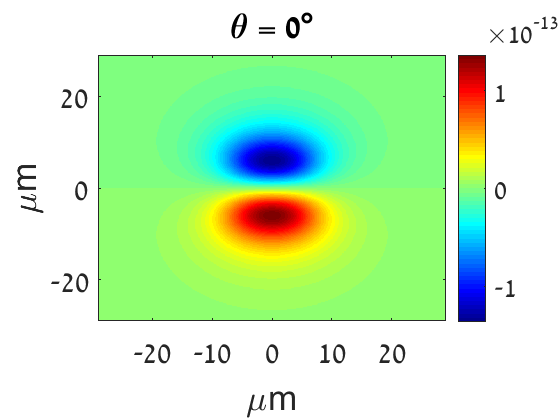}}
\subfigure[]
{\includegraphics[width = 0.32 \linewidth]{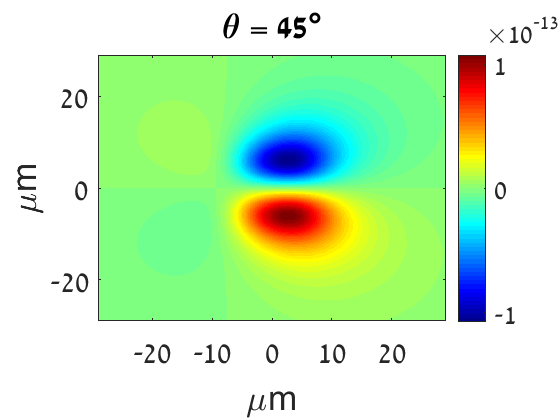}}
\subfigure[]
{\includegraphics[width = 0.32 \linewidth]{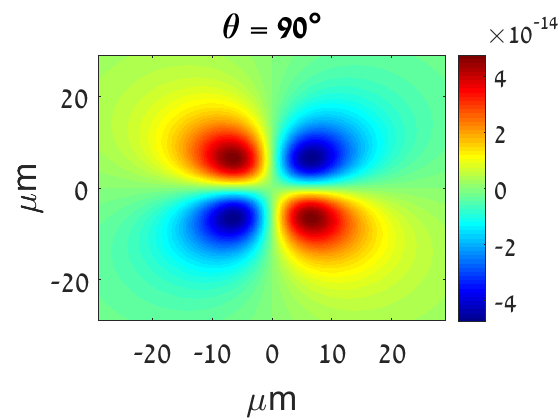}}
\caption{Simulation of the vectorial magnetization (Arb. units) in the Y direction as a function of dipole angle $\theta$, from 0 (perpendicular to sample plane) to 90 degrees from a distance of 7 $\mu$m from the sample. Azimuth angle $\phi$ = 0}. 
\label{fig:analysis_theta_y}
\end{figure}

The analysis of the stray magnetic field images measured by the NV magnetic microscope rely on the dipole approximation, as described in the main text.
\newline Here we detail our analysis of the measured widefield images of the vectorial stray field with respect to the orientation of the effective magnetic dipole of the sample's magnetization, described by polar (tilt) angle (angle with respect to the axis normal to the surface, Z) "$\theta$" and azimuthal angle (angle of rotation around Z axis, from an arbitrarily defined X axis) "$\phi$". 
\newline We first simulate a single dipole and vary its tilt angle $\theta$ with respect to the surface normal, assuming an azimuthal angle $\phi=0$. The resulting vectorial magnetic field images along the Z, X and Y directions is given in figures \ref{fig:analysis_theta_z}, \ref{fig:analysis_theta_x} and \ref{fig:analysis_theta_y}, respectively, for $\theta=0, 45^{\degree}, 90^{\degree}$, with a standoff distance of $7 \mu$m.

It is clear that the magnetic field map changes significantly as $\theta$ increases. Shifting the dipole direction towards the plane creates a second "blob" in the stray field along the Z direction, gaining magnitude as $\theta$ increases. Along the X and Y directions, a second pair of "blobs" appears in the Y direction and a third "blob" appears in the X direction, increasing their magnitude as $\theta$ grows. 



\begin{figure}[hbt!]
\subfigure[]
{\includegraphics[width = 0.32 \linewidth]{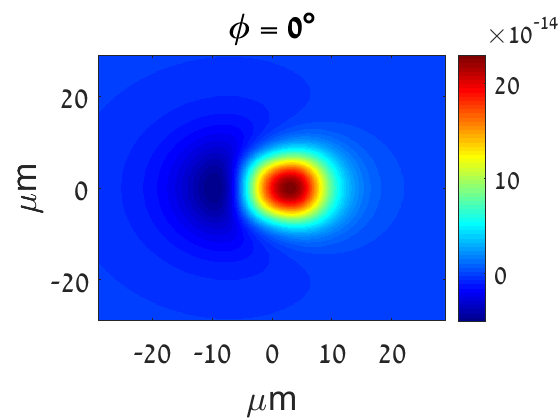}}
\subfigure[]
{\includegraphics[width = 0.32 \linewidth]{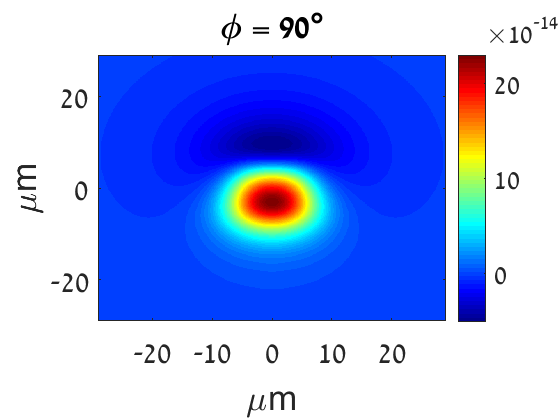}}
\subfigure[]
{\includegraphics[width = 0.32 \linewidth]{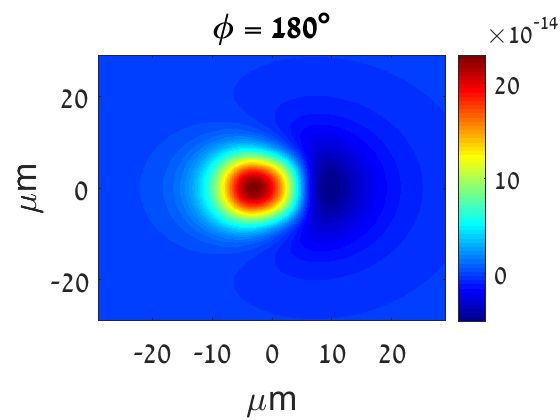}}

\caption{Simulation of the vectorial magnetization (Arb. units) in the Z direction as a function of dipole angle $\phi$, from 0 (perpendicular to sample plane) to 180 degrees from a distance of 7 $\mu m$ from the sample. The altitude angle $\theta$ is 45 degrees}.
\label{fig:analysis_phi_z}
\end{figure}

\begin{figure}[hbt!]
\subfigure[]
{\includegraphics[width = 0.32 \linewidth]{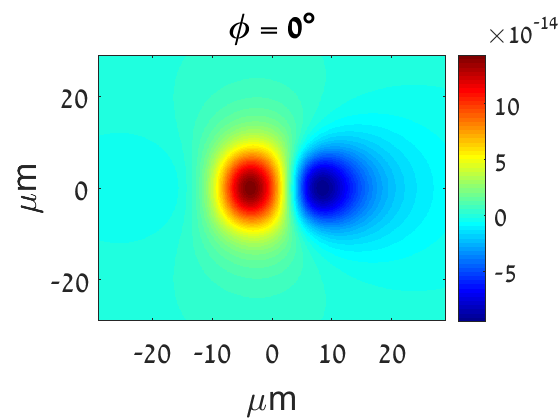}}
\subfigure[]
{\includegraphics[width = 0.32 \linewidth]{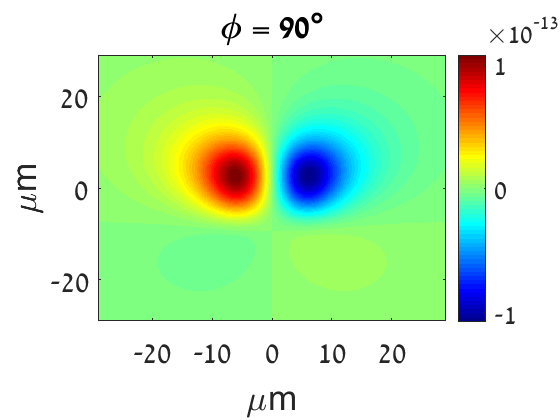}}
\subfigure[]
{\includegraphics[width = 0.32 \linewidth]{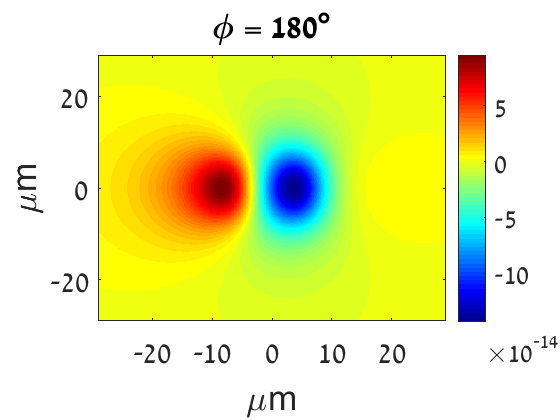}}

\caption{Simulation of the vectorial magnetization (Arb. units) in the X direction as a function of dipole angle $\phi$, from 0 (perpendicular to sample plane) to 180 degrees from a distance of 7 $\mu m$ from the sample. The altitude angle $\theta$ is 45 degrees}.
\label{fig:analysis_phi_x}
\end{figure}

\begin{figure}[hbt!]
\subfigure[]
{\includegraphics[width = 0.32 \linewidth]{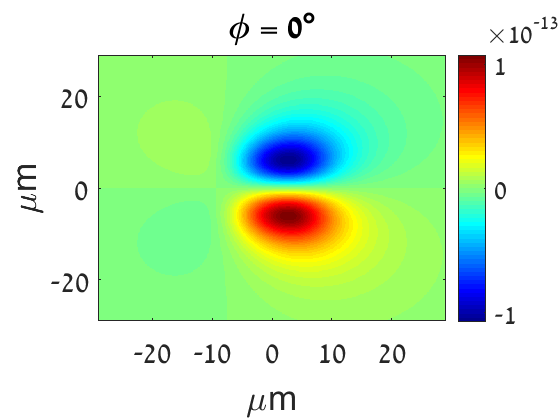}}
\subfigure[]
{\includegraphics[width = 0.32 \linewidth]{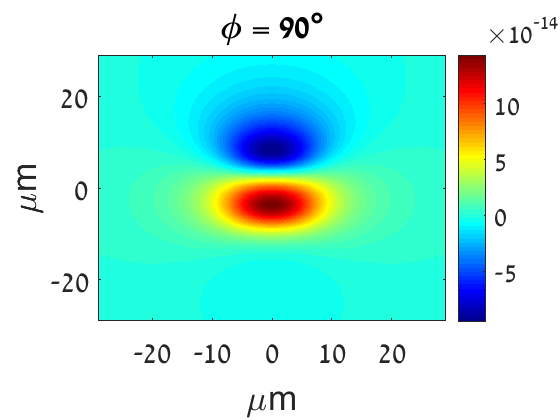}}
\subfigure[]
{\includegraphics[width = 0.32 \linewidth]{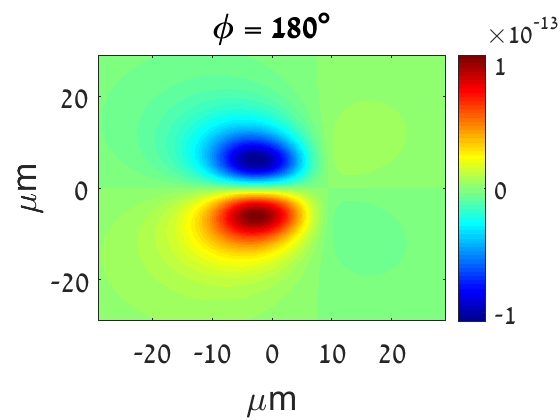}}

\caption{Simulation of the vectorial magnetization (Arb. units) in the Y direction as a function of dipole angle $\phi$, from 0 (perpendicular to sample plane) to 180 degrees from a distance of 7 $\mu m$ from the sample. The altitude angle $\theta$ is 45 degrees}.
\label{fig:analysis_phi_y}
\end{figure}

Next, we incorporate an azimuthal angle $\phi$, while maintaining a constant altitude angle $\theta$ of 45 degrees, for Z (see figure \ref{fig:analysis_phi_z}), as well as X and Y (figures \ref{fig:analysis_phi_x}, \ref{fig:analysis_phi_y}). 

\begin{figure}[hbt!]
\subfigure[]
 {\includegraphics[width = 0.3 \linewidth]{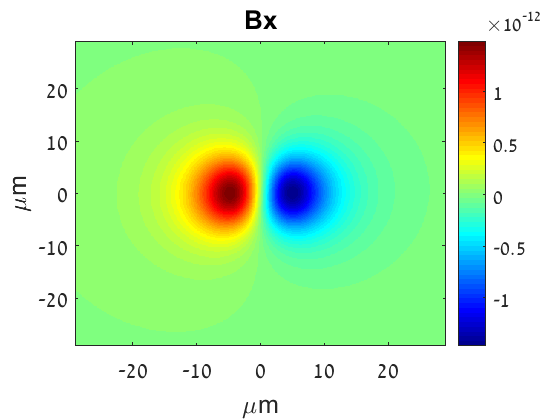}}
\subfigure[]
 {\includegraphics[width = 0.3 \linewidth]{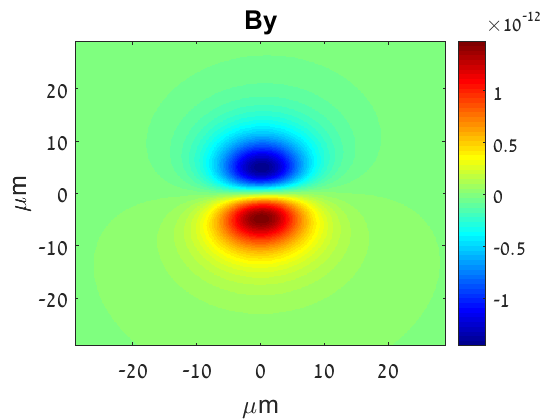}} \subfigure[]
 {\includegraphics[width = 0.3 \linewidth]{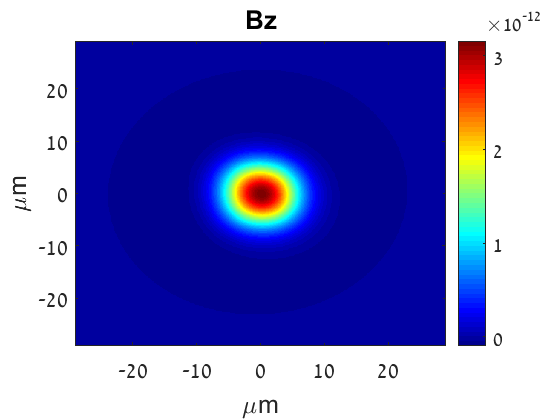}}
 \caption{Simulation of X (a), Y (b) and Z (c) components of the magnetic field from a 7 $\mu$m distance from a FM, with $\theta = 45 \degree$ random azimuth angle ($\phi$). The magnetic field images seem similar to a magnetic field generated by out-of-plane magnetization of the FM ($\theta = 0$), since the in-plane magnetization is cancelled by dipoles aligned in opposite azimuth angles}. 
\label{fig:randomPhi}
\end{figure}

The azimuthal angle of the molecules is manifested in the orientation of the primary and the secondary "blobs" in the Z component of the magnetic field. Random angle, combined with the relatively large standoff distance between the diamond and the FM sample, would eliminate the in-plane field (as opposite angle would cancel each other's in-plane component). Thus, the secondary "blob" in the Z component of the magnetic field would disappear, leaving only the primary "blob" present, while for X and Y components the two "blobs" would become perfectly symmetric, both in position and magnitude. A simulation showing the expected fields assuming random azimuthal angle is given in Fig. \ref{fig:randomPhi}. 

\begin{figure}[hbt!]
\subfigure[]
{\includegraphics[width = 0.45 \linewidth]{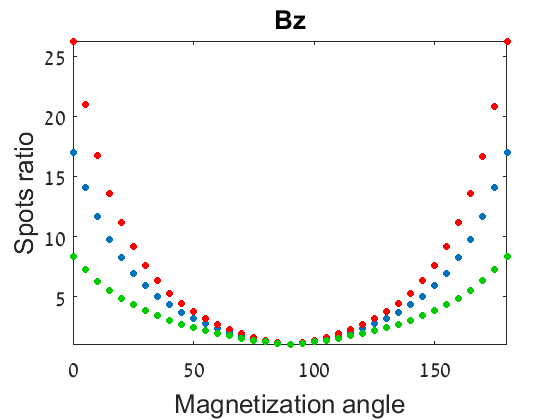}}
\subfigure[]
{\includegraphics[width = 0.45 \linewidth]{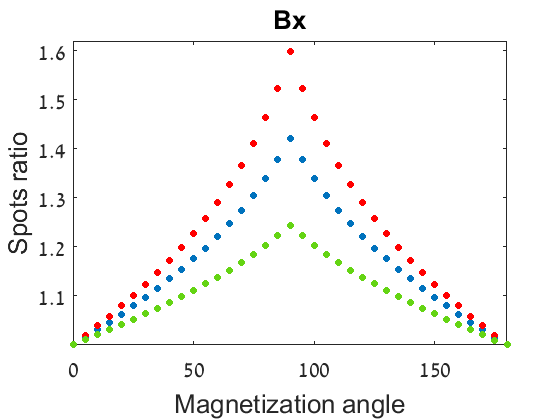}}
\caption{The ratio between the two main "blobs" in Bz (a) and Bx (b), as a function of magnetization angle and standoff distance between the diamond and the ferromagnetic sample. Green dots - 3 $\mu$m, blue dots - 7 $\mu$m, red dots - 11 $\mu$m.} 
\label{fig:blob_ratio}
\end{figure}

The distance between the FM sample and the diamond slightly modifies the ratio between the blob  magnitudes. Figure \ref{fig:blob_ratio} depicts the "blob" amplitude ratio range as a function of $\theta$ for Bx and Bz, for 3 different heights from the sample - 3 $\mu m$, 7 $\mu m$ and 11 $\mu m$. Although the ratios change for different standoff distances between the diamond and the ferromagnet, they still provide a decent estimation for the magnetization angle, even for such a large error in the standoff distance. As seen from the figures, analyzing the ratio of the "blobs" in the X direction provides us with a good resolution for small angles, while the Z direction is better for angles close to 90 degrees. 

\section{Magnetic characterization of the FM thin film}
\begin{figure}[hbt!]
 {\includegraphics[width = 0.8 \linewidth]{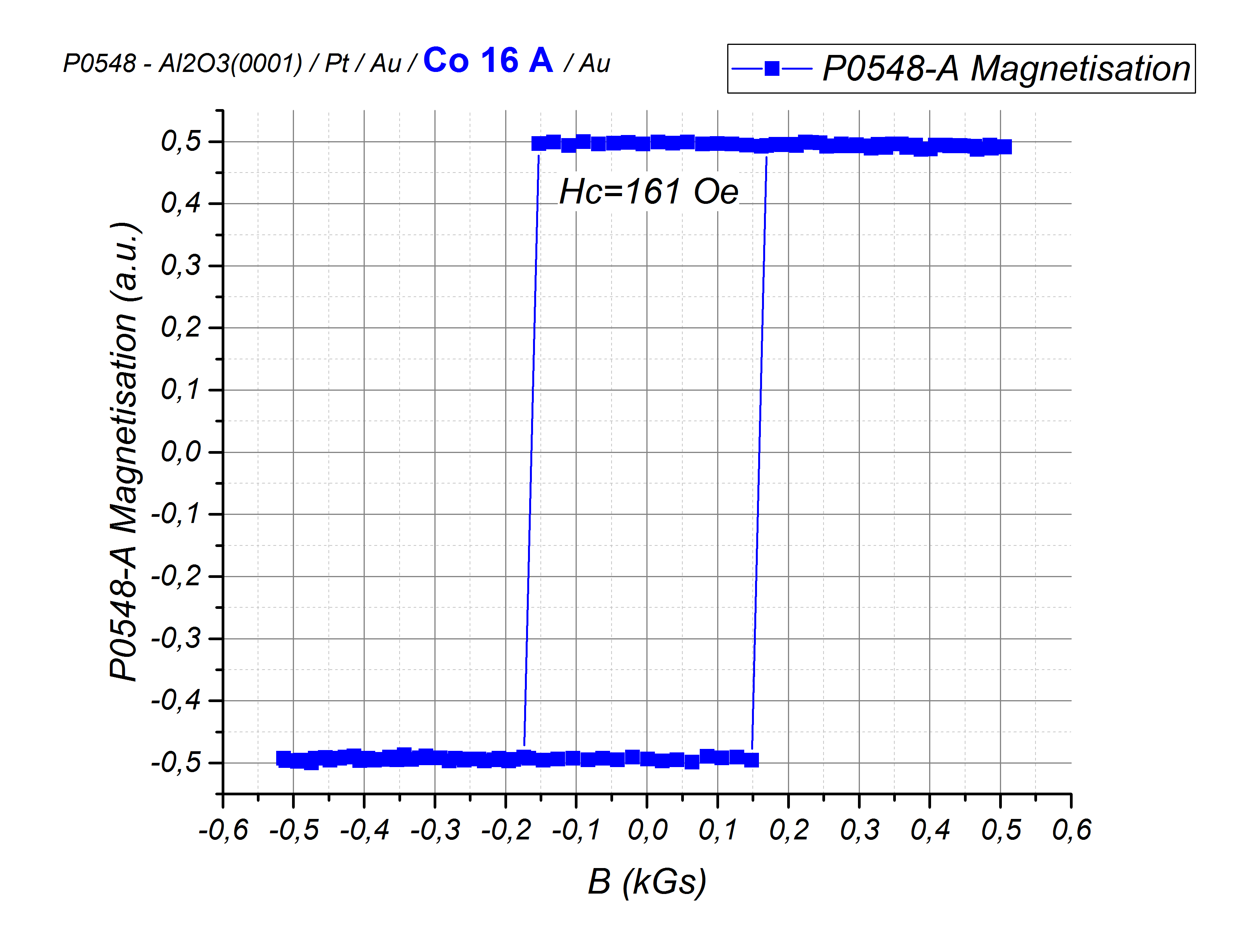}}
 \caption{Hysteresis loop measured using P-MOKE. The measurement was done on a 1.6nm thickness of Co, showing a strong perpendicular anisotropy, coercive field of 161G and remanence value close to 1. } 
\label{fig:MOKE}
\end{figure}

The magnetic properties of all FM thin film samples were probed by loop using polar magneto optic Kerr (P-MOKE) effect to measure a hysteresis loop. Figure \ref{fig:MOKE} exhibits an exemplary measurement. This hysteresis loop provides evidence for a strong perpendicular anisotropy, coercive field of $\sim$160G and a remanence value close to 1.   

\section{Alpha Helix L-Polyalanine specifications}
\begin{figure}[hbt!]
 {\includegraphics[width = 0.55 \linewidth]{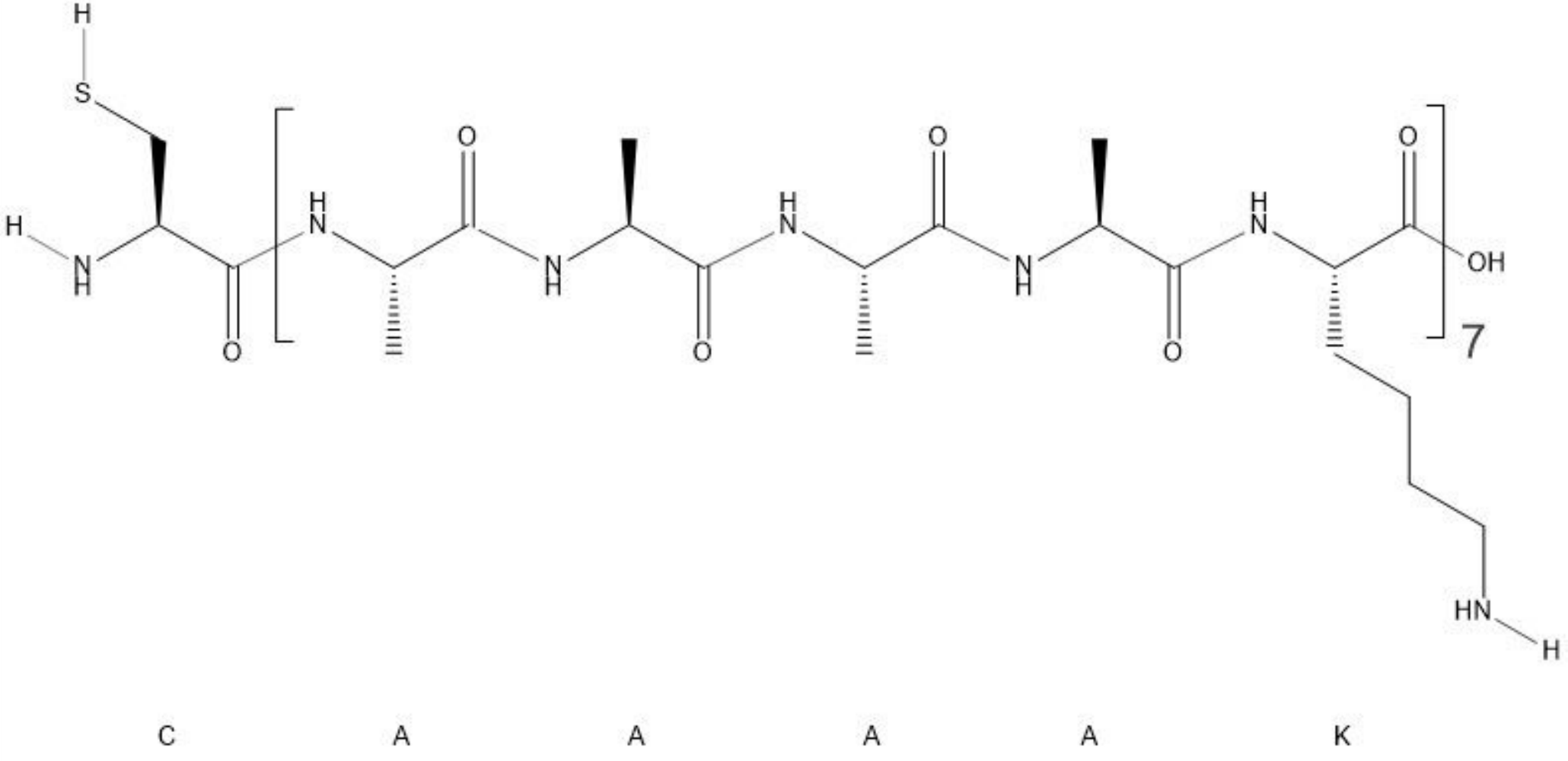}}
 \caption{Schematic of the molecules used in this research.
Alpha helix L-polyalanine with the following sequence: \newline
 [H]-CAAAAKAAAAKAAAAKAAAAKAAAAKAAAAKAAAAK-[OH] 
\\The theoretical length of the molecule is 5.4nm.} 
\label{fig:AHPA}
\end{figure}

The chemical structure of the AHPA used in this work is specified in Figure \ref{fig:AHPA}. These molecules have a helix structure and theoretical length of $\sim$5.4nm.

\subsection{FTIR measurements}
\begin{figure}[hbt!]
 {\includegraphics[width = 0.6 \linewidth]{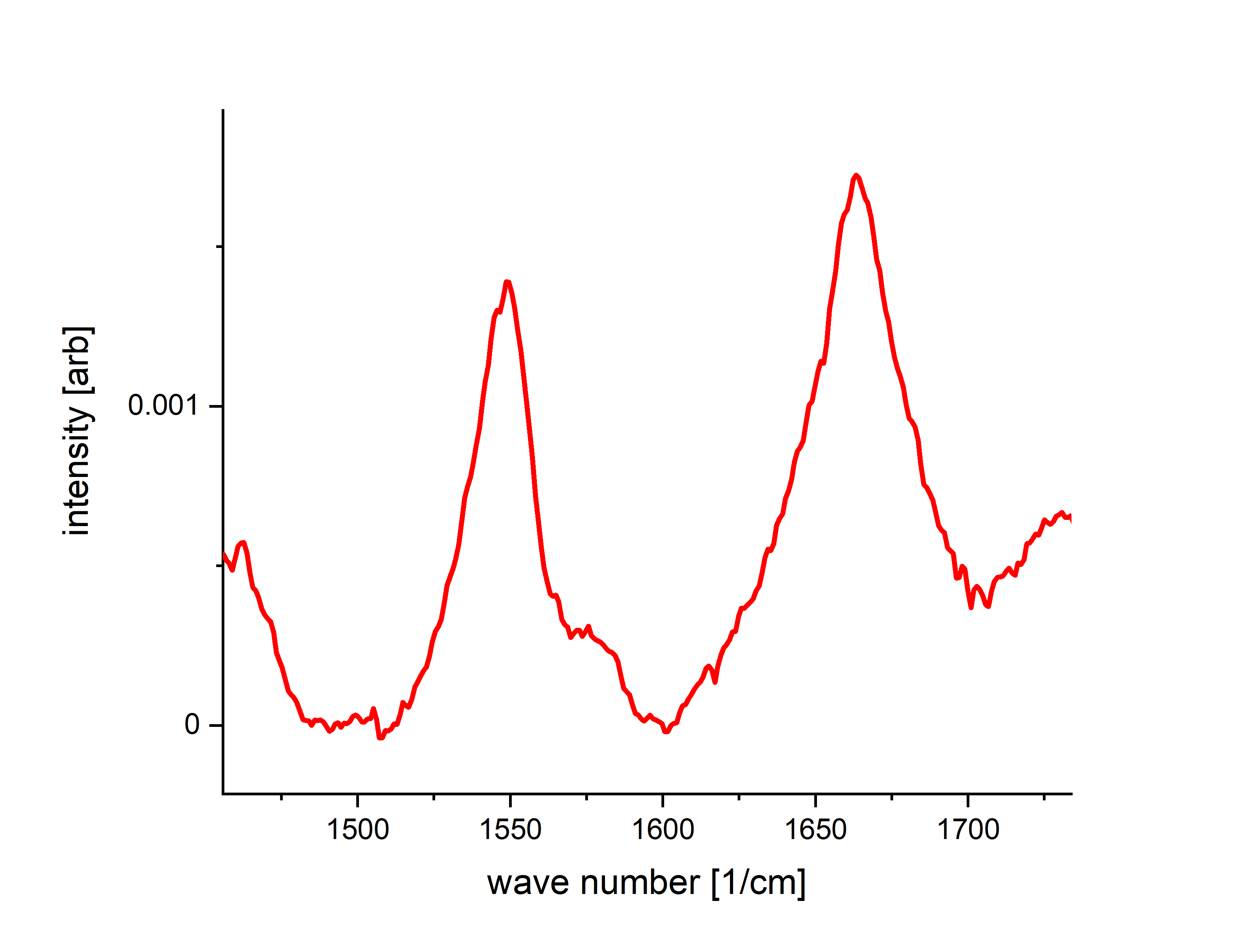}}
 \caption{Polarization modulation–infrared reflection–absorption mode in FTIR measurement of the tilt angle of the AHPA molecules. Analysis of the ratio between peaks as done elsewhere\cite{tassinari_chirality_2018}, give a tilt angle of $\sim$60$\degree$, similar to the tilt angles measured using AFM.} 
\label{fig:FTIR}
\end{figure}

Figure \ref{fig:FTIR} shows results of the molecular monolayer tilt angle measured by Polarization modulation - infrared reflection absorption mode in FTIR (PM-IRRAS). The PM-IRRAS spectra were measured on a Nicolet 6700 FTIR instrument equipped with a PEM-90 photoelastic modulator (Hinds Instruments, Hillsboro, OR) at an 80$\degree$ angle of incidence. The orientation of the peptide monolayer on the substrate was calculated using the equation in Ref. \cite{tassinari_chirality_2018} and yielded an angle of $\sim$60$\degree$. 

\section{AFM height analysis}
Large number of randomly selected cross-sections were analyzed by fitting to two Gaussian functions. The height difference was then deduced by the distance between the centers of the two Gaussians. All different heights from each cross-sections were averaged and the standard deviation was taken as the error. 

\section{Author Contributions}
$^\nabla$ I.M., N.S., and G.H. contributed equally to this work.

\bibliography{ChiralMolecules.bib}